%% file: VisMemo.tex
\documentclass[sigplan,twocolumn]{acmart}

\usepackage{booktabs}
\usepackage{array}
\usepackage{enumitem}
\usepackage{textgreek}
\usepackage{subcaption}
\usepackage{algorithmicx}
\usepackage{algorithm}
\usepackage{algcompatible}
\usepackage{subcaption} 
\usepackage{tikz}
\usepackage{adjustbox}
\usepackage{threeparttable}
\usepackage{soul}
\sethlcolor{yellow}
\usepackage{xcolor}

\setcopyright{none}
\renewcommand\footnotetextcopyrightpermission[1]{}
\definecolor{annotationdarkred}{RGB}{139,0,0}

\definecolor{annotationdarkgreen}{RGB}{0,100,0}

\newcommand{\system}{VisMemo}

\newcommand*\circled[1]{\tikz[baseline=(char.base)]{
    \node[shape=circle,draw,fill=white,text=black,inner sep=1pt] (char) {#1};}}

\begin{document}

\title{Visual Cache Reuse for Efficient Multimodal Inference}

\makeatletter
\renewcommand{\@authorfont}{\normalsize\normalfont}
\makeatother
\author{
  Shuheng Gao$^{1,2*\ddagger}$,
  Yuankai Fan$^{2*}$,
  Jiaqiang Ma$^{2,3}$,
  Jiawei Shao$^{2}$,
  Yongkun Li$^{1\dagger}$,
  Xuelong Li$^{2\dagger}$
}
\affiliation{
  \institution{\small
    \textsuperscript{1}University of Science and Technology of China\quad
    \textsuperscript{2}Institute of Artificial Intelligence (TeleAI), China Telecom\\
    \textsuperscript{3}National University of Defense Technology}
    \country{}
}

\begin{abstract}
Vision-language model (VLM) inference incurs substantial prefill latency and memory overhead due to the long visual token sequences of large visual inputs. Existing serving systems typically exploit exact cache reuse through prefix caching or position-independent caching, overlooking the potential of approximate caching in real-world workloads. However, directly reusing an approximate cache entry may introduce spurious visual information, which na\"{i}ve approaches that select a subset of important visual tokens for recomputation fail to adequately mitigate this risk.

We present \system, an efficient serving system that enables approximate KV cache reuse for modern VLMs. Its core is a \emph{system-algorithm co-design} that combines an entropy-aware KV cache reuse algorithm with differentiated cache management. Specifically, \system\ uses patch entropy to identify reuse candidates and warmup attention to prioritize recomputation based on the current query. To support the large cache capacity across GPU memory, CPU memory, and SSD storage, \system\ applies distinct I/O and cache policies to exact and approximate hits: sparse patch-level I/O for approximate hits, prioritized retention for exact hits, and a reserved CPU victim cache for preset entries selected using offline workload priors and online reuse observations. \system\ further optimizes the memory access path between the inference engine and the storage backend, eliminating redundant caching and unnecessary data movement. Experiments on multiple public datasets show that \system\ preserves output quality under approximate reuse. On a real-world production trace, \system\ increases the overall cache hit rate from 27.78\% to 61.45\% and reduces prefill latency by up to 48.3\% at 8K input tokens.
\end{abstract}

\maketitle
\begingroup
\renewcommand\thefootnote{}
\footnotetext{$^{*}$Equal contribution.}
\footnotetext{$^{\dagger}$Corresponding authors.}
\footnotetext{$^{\ddagger}$Work done during internship at TeleAI.}
\endgroup
\pagestyle{plain}
\thispagestyle{plain}

\section{Introduction}
Vision-language models (VLMs) are widely deployed in interactive real-time applications such as visual question answering (VQA)~\cite{wang2024qwen2vl,mathew2021docvqa,mathew2022infographicvqa}, UI agent automation~\cite{xie2024osworld,bai2025qwen3vl,vteam2025glm}, and multimodal search~\cite{faysse2025colpali,jiang2025vlm2vec,lin2025mmembed,an2026aiflow}, where a low time-to-first-token (TTFT) is required to provide a responsive user experience~\cite{zhong2024distserve,qiu2025modserve}. In these models, high-resolution images and videos are first encoded into long sequences of visual tokens~\cite{cao2023pumer}, which are then concatenated with the text tokens~\cite{chen2024fastv,yang2025visionzip,fu2025framefusion}. During prefilling, the resulting attention computation grows quadratically with input sequence length~\cite{vaswani2017attention}, making visual inputs a major source of latency while their key-value (KV) cache occupies substantial GPU memory and strains both computation and memory resources~\cite{kwon2023pagedattention,tu2024vlcachecompression,wan2025meda,pei2024crossself}.

\begin{figure}[t]
    \centering
    \begin{subfigure}{\linewidth}
        \centering
        \includegraphics[width=\linewidth]{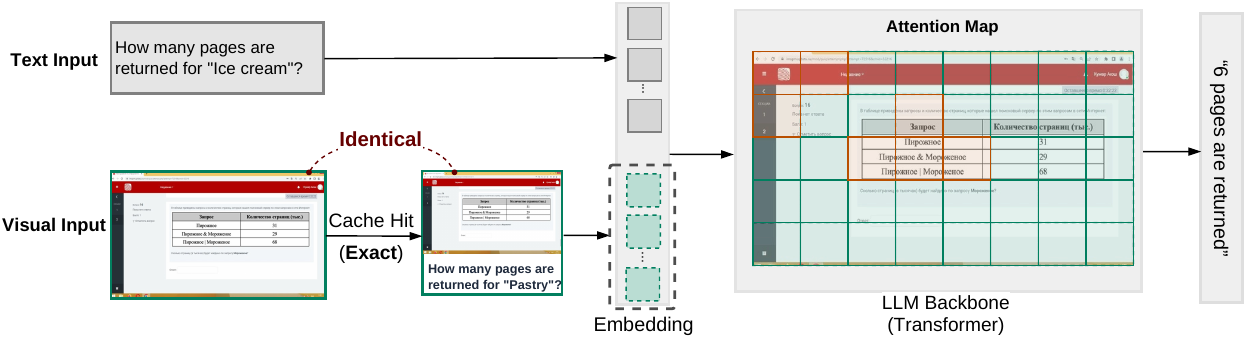}
        \caption{Position-independent KV caching.}
        \label{fig:cache_reuse_flow_a}
    \end{subfigure}
    \begin{subfigure}{\linewidth}
        \centering
        \includegraphics[width=\linewidth]{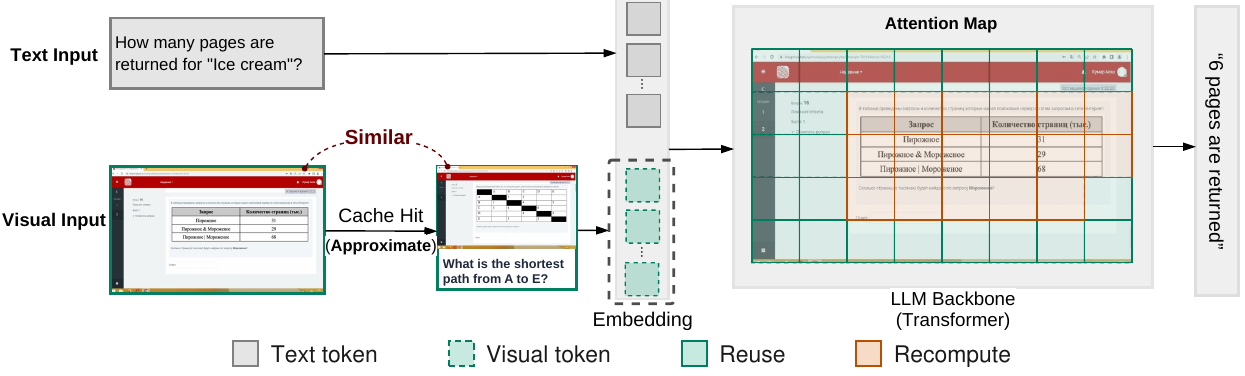}
        \caption{Approximate KV caching.}
        \label{fig:cache_reuse_flow_b}
    \end{subfigure}
    \caption{Illustration of two KV cache reuse strategies.}
    \label{fig:cache_reuse_flow}
\end{figure}

To accelerate VLM inference, as with large language models (LLMs), existing serving systems commonly employ prefix caching, reusing KV caches across requests that share the same prefix~\cite{gim2024promptcache, ye2024chunkattention, zheng2024sglang}. Considering that the prefix-match requirement is often unmet when textual prompts vary, position-independent KV caching has emerged to expand cache-reuse opportunities by selectively recomputing cache for only a small fraction of visual tokens to account for changed positional contexts~\cite{hu2024epic, yao2025cacheblend, qin2025vlcache, ma2026kamera,zhao2025mpic} (as depicted in~\autoref{fig:cache_reuse_flow_a}). Nevertheless, these approaches still rely on \emph{exact} token-level matching and can reuse visual KV caches only when the visual token sequences are identical~\cite{qin2025vlcache,ma2026kamera}. In practice, VLM workloads often contain highly similar visual inputs, such as partially modified screenshots. Exact token-level matching therefore treats these requests as cache misses, even when most visual patches remain unchanged, leaving substantial reusable visual KV cache unexploited (see $\S\ref{2.1}$ for more details). This reveals \emph{approximate} KV reuse opportunities beyond the identical visual match.

However, approximate caching for VLMs is \textbf{non-trivial}. Visual similarity alone does not imply that the entire (visual) KV cache remains valid: an approximate match may differ in fine-grained details (e.g., a single table entry) that are critical to the query~\cite{mathew2021docvqa}. In addition, visual KV representations also depend on the preceding textual context; changes to the textual prefix can therefore alter these representations and compromise cache reuse~\cite{hu2024epic,yao2025cacheblend}. A naïve approach to approximate caching is \emph{recomputation}~\cite{chen2024fastv,fu2025framefusion}. Similar to position-independent KV caching methods, recomputation provides a naïve form of approximate caching: it recomputes a larger fraction of high-importance tokens while reusing the KV caches of low-importance counterparts (\autoref{fig:cache_reuse_flow_b}). Unfortunately, severe token sinks and attention noise make the selection unreliable for estimating the importance of visual tokens~\cite{kang2025visualattentionsink,xi2026lostattention} (more details can be seen in $\S\ref{2.2}$).

In this paper, we introduce \system, an efficient VLM inference system that leverages approximate KV cache reuse to reduce serving latency while preserving generation quality. \system\ is motivated by our observation that \emph{entropy}~\cite{shannon1948mathematical} provides a stable signal for estimating the risk of reusing visual patches\footnote{A patch is an image region encoded as a visual token. Here, entropy serves as a proxy for model uncertainty.}: low-entropy patches generally encode limited visual information and are more likely to remain stable when reusing, whereas high-entropy patches capture richer visual details and therefore require more conservative treatment. Based on this insight, we develop an \emph{entropy-aware algorithm} for approximate KV cache reuse. It selects patches that have low entropy in both images as reuse candidates, then uses the normalized attention score collected during a short warmup phase~\cite{yao2025cacheblend,qin2025vlcache} to prioritize recomputation among the remaining patches according to the current user query. By using entropy to identify reusable patches and attention to prioritize recomputation, it avoids solely relying on a noisy attention map while preserving visual information relevant to the current query ($\S\ref{3.3}$).

Although the proposed entropy-aware algorithm preserves output quality, realizing the benefits of approximate caching requires accommodating a substantially larger KV cache footprint. This, in turn, necessitates the system-level optimization atop existing hierarchical KV cache architectures~\cite{kwon2023pagedattention,yang2025kvshare,qin2025mooncake,liu2025lmcache,feng2025adaptcache} that incorporates heterogeneous storage tiers while maintaining low-latency performance. First, a unified system must support two cache-reuse modes with distinct access patterns: exact reuse retrieves an entire KV cache entry and exhibits strong locality, whereas approximate reuse retrieves only the KV caches of selected patches and exhibits weaker locality. To accommodate these differences, we propose a \emph{differentiated cache-management mechanism} that efficiently supports both reuse modes within a hierarchical KV cache architecture. Second, we introduce a \emph{prior-guided placement strategy} that combines offline workload priors with online reuse signals to determine which cache entries to retain in a dedicated CPU-side victim cache ($\S\ref{3.4}$).

To evaluate the performance of \system\, we conduct comprehensive experiments on a real-world conversational VLM workload~\cite{chou2025visionarena} and four public visual understanding benchmarks~\cite{mathew2022infographicvqa, mathew2021docvqa, liu2024mmbench, li2023seedbench}, using the Qwen3-VL~\cite{bai2025qwen3vl} and GLM-4.6V~\cite{vteam2025glm} models. The experiment results show that \system\ outperforms existing solutions in approximate-hit quality, improving ANLS by up to \textbf{19.72} points over the stronger baseline while maintaining comparable exact-hit quality. Notably, approximate matching increases the overall cache-hit rate from 27.23\% to \textbf{60.28\%}, while approximate reuse reduces prefill latency by up to \textbf{48.3\%} compared with full prefill on the real-world workload. Our system-level optimizations further accelerate both exact and approximate hits relative to the original Mooncake backend~\cite{qin2025mooncake}.

We summarize our main contributions as follows:
\begin{itemize}[leftmargin=6pt]
\item \textbf{Approximate KV cache reuse for VLMs}. We identify substantial opportunities for KV cache reuse beyond exact visual matches in real-world VLM workloads, to the best of our knowledge, are the first to introduce \emph{approximate caching} for efficient VLM inference.

\item \textbf{System-algorithm co-design}. We propose \system, an approximate caching system for VLM inference that combines an efficient KV cache reuse algorithm with effective hierarchical cache management. At the algorithmic level, \system\ introduces an entropy-aware method that selects high-relevant patches for recomputation and low-entropy patches for reuse. At the system level, \system\ designs a differentiated cache management scheme to support both exact and approximate reuse via sparse I/O at patch-granular, coupled with prioritized retention and eviction policies.

\item\textbf{Practical system implementation and evaluation}. We implement \system\ by extending SGLang~\cite{zheng2024sglang} and Mooncake. \system\ achieves the highest overall scores in the five datasets, increases the total cache hit rate from 27.78\% to 61.45\%, and reduces prefill latency by up to 48.3\%.
\end{itemize}

\section{Background and Motivation}
\subsection{VLM Inference with Visual KV Cache Reuse}\label{2.1}
\begin{figure}[t]
\centering
\includegraphics[width=\linewidth]{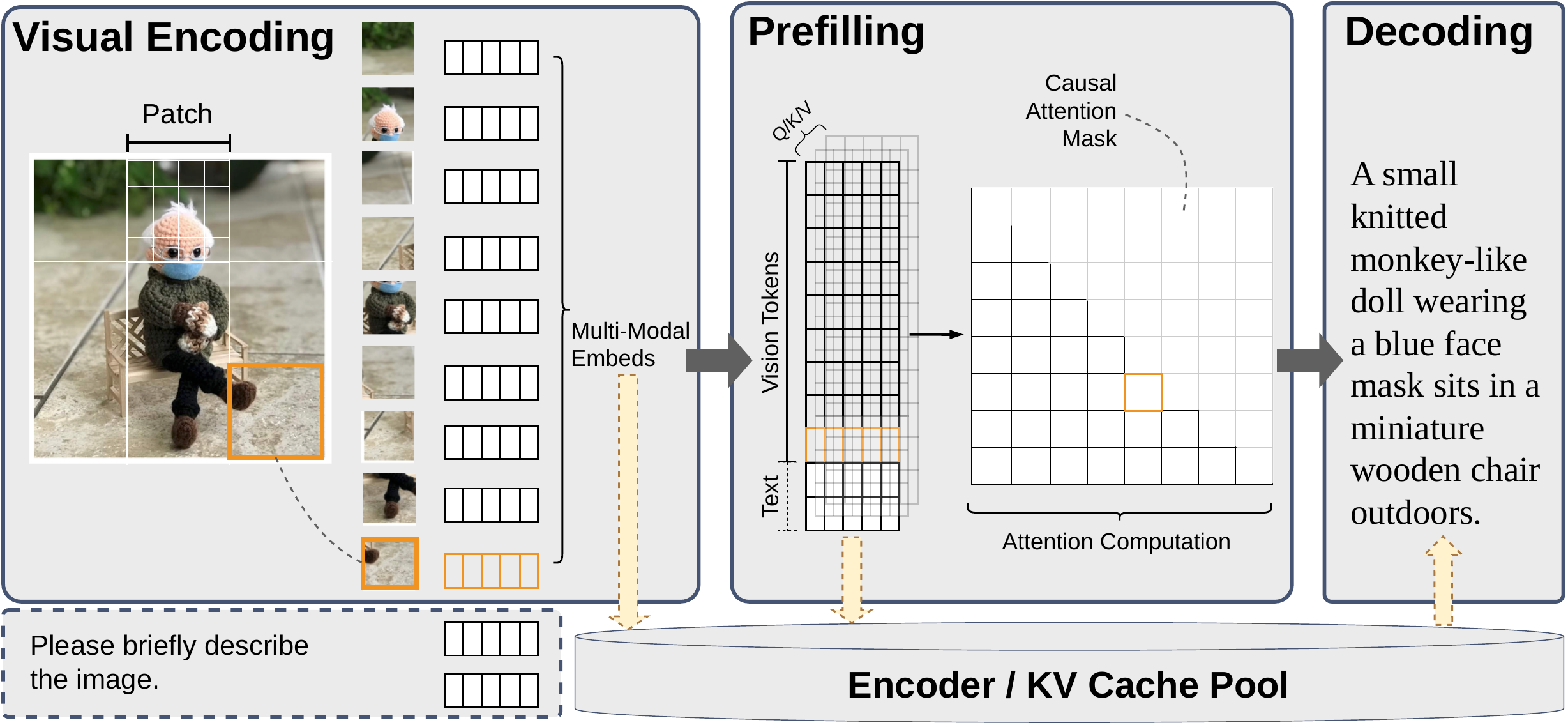}
\caption{Three-stage VLM inference process, including visual encoding, prefilling, and decoding.}
\label{fig:epd}
\end{figure}

VLM inference extends the autoregressive LLM paradigm by incorporating a visual encoder (e.g., Vision Transformers for images/video) to support multimodal understanding~\cite{wang2024qwen2vl, bai2025qwen3vl, vasu2025fastvlm}. It proceeds in three stages: visual encoding, prefilling, and decoding~\cite{singh2025epd}, as illustrated in~\autoref{fig:epd}. (1) During visual encoding, the image is divided into patches and transformed by the visual encoder into a sequence of visual tokens (i.e., embedding tensors). These visual tokens are stored as the \emph{encoder cache}, indexed by a multimodal hash, and concatenated with the text tokens for prefilling. Here, a single image may produce thousands of visual tokens, depending on its resolution~\cite{chen2024fastv, yang2025visionzip, arif2025hired, zhang2025sparsevlm, bolya2023tome,shang2025llavaprumerge}, often far exceeding the number of text tokens. (2) During prefilling, each transformer layer projects the hidden state into query, key, and value vectors for every input token. Self-attention computes weights from queries and keys to aggregate the values; the resulting key and value vectors are retained as the \emph{KV cache}, while the attention output is passed through the feed-forward network (FFN) to produce hidden states in the next layer~\cite{vaswani2017attention}. (3) During decoding, the model generates one token at a time and reads the KV cache rather than recomputing prior tokens~\cite{kwon2023pagedattention}.

\begin{figure}[t]
\centering
\begin{minipage}{0.49\textwidth}
\centering
\begin{subfigure}{0.49\linewidth}
    \includegraphics[width=\linewidth]{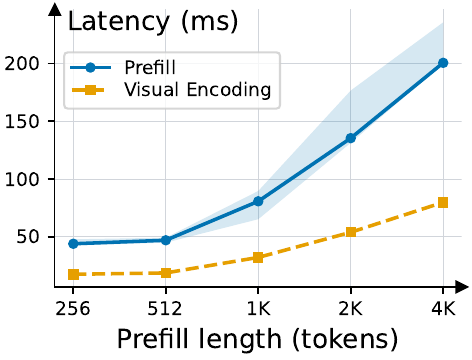}
    \caption{Encoding and prefill latency}
    \label{fig:prefill_embedding_latency}
\end{subfigure}
\hfill
\begin{subfigure}{0.49\linewidth}
    \includegraphics[width=\linewidth]{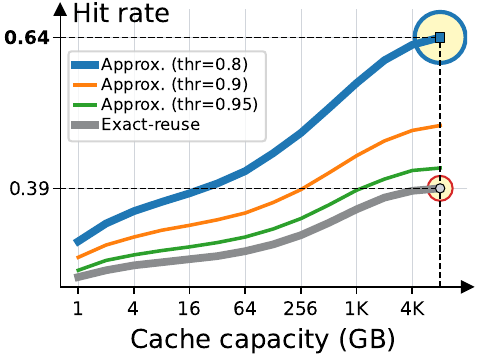}
    \caption{Cache hit rate vs. capacity}
    \label{fig:hit_rate_vs_capacity}
\end{subfigure}
\end{minipage}
\caption{Two motivating observations: (a) Prefill latency increasingly dominates with sequence length; (b) Scaling trend of cache hit rates with increasing storage capacity.}
\label{fig:reuse_motivation}
\end{figure}

It's worth noting that incorporating visual tokens into the input sequence substantially increases both the computational and memory requirements during prefill~\cite{zhong2024distserve, agrawal2024sarathiserve, zheng2024sglang, fan2026tradeoffs}. First, attention complexity grows rapidly with sequence length, requiring every transformer layer to process the full visual token sequence before generating the first output token. Second, the visual KV cache imposes a substantial memory burden with its footprint growing linearly with the sequence length~\cite{liu2024cachegen, qin2025mooncake, tu2024vlcachecompression, liu2024kivi, wan2024lookm,li2025madakv}. This growth quickly exhausts available memory and limits long-horizon inference~\cite{zhang2023h2o, li2024snapkv, tang2024quest,chen2026streamkv,lee2024infinigen,liu2025retrievalattention}. These costs are reflected in~\autoref{fig:prefill_embedding_latency}: for Qwen3-VL-32B~\cite{bai2025qwen3vl} on two NVIDIA H100 GPUs, prefill latency increases sharply with input length and reaches $\backsim$2.5$\times$ the latency of vision encoding. Visual prefill therefore constitutes a major computational and memory bottleneck in VLM serving.

To reduce visual prefill overhead, serving systems commonly employ prefix caching~\cite{gim2024promptcache, ye2024chunkattention, yang2025kvshare}, which reuses the KV cache when requests share an identical prefix. This requirement is restrictive: even identical visual inputs (e.g., images) incur cache misses when their surrounding textual prefixes differ~\cite{hu2024epic, ma2026kamera}. Representative systems such as VLCache~\cite{qin2025vlcache} and CacheBlend~\cite{yao2025cacheblend} address this limitation through \emph{position-independent KV cache reuse}, retrieving cached image representations whenever an image recurs at the prefix or elsewhere in the sequence. Because directly reusing the visual KV cache skips cross-attention from visual tokens to the current textual prefix, the systems selectively recompute the KV cache of critical visual tokens~\cite{qin2025vlcache, yao2025cacheblend}. At each transformer layer, selected tokens perform attention and FFN computations conditioned on the current prefix, while the remaining tokens retain their cached states. Specifically, CacheBlend identifies the high deviation tokens by comparing cached and newly computed KV cache after a lightweight first-layer warm-up, then progressively reduces the recomputation set across layers~\cite{yao2025cacheblend}. VLCache instead uses offline layer-sensitivity profiling to determine per-layer recomputation ratios and prioritizes early visual tokens, whose reuse errors may propagate to subsequent tokens~\cite{qin2025vlcache}. Both systems focus on the most effective token computation to minimize reuse errors.

\subsection{Approximate Reuse Opportunities}\label{2.2}
Position-independent caching allows an identical visual input to reuse its KV cache across different input positions~\cite{liu2026qvpic}, but still requires an exact match. This constraint misses common reuse opportunities in real-world VLM workloads, where visual inputs often undergo \emph{localized modifications}. 
For example, UI screenshots may retain navigation bars, panels, and backgrounds while changing only a widget or status indicator~\cite{xie2024osworld}; document revisions may preserve page templates and table structures while altering limited content~\cite{chou2025visionarena}; and consecutive video frames typically retain most of the scene while modifying a foreground object or local state~\cite{fu2025framefusion}. 
Although such inputs are assigned distinct multimodal hash keys, their visual content often remains largely unchanged, leading us to posit that this shared content can be approximately reused.
\autoref{fig:hit_rate_vs_capacity} quantifies this approximate-reuse opportunity by sampling 10,000 requests from VisionArena~\cite{chou2025visionarena} and using Qwen3-VL-32B with a LRU eviction strategy: under an approximate-reuse criteria with a semantic-similarity threshold of 0.8, the system increases the cache hit rate from 35.6\% to 56.1\% with a 1024~GB cache and further reaches 63.5\% with a larger capacity (8192~GB). 
These results reveal a \textbf{scaling trend} between cache capacity and approximate-reuse hit rate, revealing substantial reuse opportunities missed by exact hash-based matching.

These observations place approximate reuse between exact hits and complete misses. The unchanged regions may preserve substantial visual prefill computation, but the modified regions prevent the cached entry from being treated as uniformly valid. Global similarity alone cannot resolve this tension because a large unchanged region may dominate the similarity score and obscure a small modification that is critical to the current prompt. Approximate cache reuse thus offers substantial potential to expand reuse, but together with an unresolved validity question: \emph{how reliable is reuse when only part of the visual content changes?}

\begin{figure}[t]
\centering
\includegraphics[width=\linewidth]{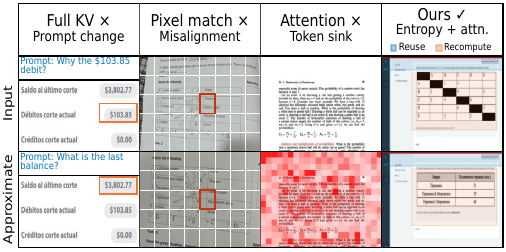}
\caption{Representative cases motivating \system\ that combines entropy and attention to distinguish reusable patches from patches requiring recomputation.}
\label{fig:approximate_reuse_cases}
\end{figure}

\subsection{Challenges of Approximate KV Cache Reuse }
Approximate caching retrieves a candidate KV entry, selecting which portions of the entry to reuse without compromising quality remains challenging. We identify three challenges by examining three intuitive strategies and their limitations, as illustrated in~\autoref{fig:approximate_reuse_cases}. 

\noindent\textbf{Challenge 1: Textual-prefix changes cause unreliable cross-attention} (\emph{1st column,~\autoref{fig:approximate_reuse_cases}}). A straightforward approach is \emph{na\"{i}ve full KV reuse}, which directly reuses the entire visual KV cache of a matched image. However, full KV reuse neither computes cross-attention from visual tokens to a changed textual prefix nor updates the KV cache for modified image regions. 

As shown in the first column of \autoref{fig:approximate_reuse_cases}, a nearly identical bank statement is queried for different information: the last balance in the cached request and the reason for the \$103.85 debit in the current request. When the changed prompt precedes the visual tokens, direct reuse will miss cross-attention from visual tokens to the current textual prefix~\cite{qin2025vlcache,yao2025cacheblend}. Reusing the whole entry also retains the KV cache associated with the modified image regions. 

\noindent\textbf{Challenge 2: Patch misalignment} \textbf{causes false-negative reuse decisions} (\emph{2nd column,~\autoref{fig:approximate_reuse_cases}}). 
Another na\"{i}ve approach is \emph{patch-based matching}, which selects the KV cache corresponding to the matched patches for approximate reuse. That is, it compares the pixel values at the same grid positions in the current and a similar image, treating patches with identical or sufficiently similar pixels as reuse candidates. This approach assumes that shared visual content remains aligned with the patch grid; however, a slight displacement of the entire image may shift unchanged content across patch boundaries, causing patch-based matching to fail. 

As shown in the second column of~\autoref{fig:approximate_reuse_cases}, changes in camera viewpoint cause the same visual content to fall into different patches. The highlighted patches (red boxes) cover different regions of the table. Thus, even when the table content remains unchanged, a positional shift can lead patch-based matching to produce false-negatives~\cite{edstedt2024roma,tang2023dift}.

\noindent\textbf{Challenge 3: Visual-token sinks cause misguided recomputation} (\emph{3rd column,~\autoref{fig:approximate_reuse_cases})}.
Intuitively, reusing the KV cache from a similar request may retain states that do not accurately reflect the current request. \emph{Recomputation}~\cite{yao2025cacheblend} addresses this issue by recomputing the visual tokens most relevant to the current prompt (while reusing the remaining cached tokens), and the attention mechanism provides an intuitive signal for selecting these tokens: those receiving higher attention scores are recomputed, whereas the others are reused.
However, attention scores alone can be misleading; some visual tokens attract a disproportionate amount of attention despite contributing little relevant information to the current prompt~\cite{kang2025visualattentionsink,xi2026lostattention}. This phenomenon, termed as \emph{visual-token sinks}~\cite{kang2025visualattentionsink}, may cause such tokens to be incorrectly prioritized for recomputation.

The third column of~\autoref{fig:approximate_reuse_cases} shows a document page (top) and its visual-token attention map (bottom), where darker red denotes higher attention to the corresponding image patches. Notably, several patches in the blank top margin receive high attention scores despite containing little visual information. Thus, an attention-only policy may ``mistakenly'' select these patches for recomputation, wasting part of the budget on low-information regions and reducing the budget available for content relevant to the current prompt.

These observations motivate separating patch-level reuse risk from relevance to the user input. \system\ uses patch entropy to identify low-risk reuse candidates and warmup attention to guide selective recomputation. The fourth column of~\autoref{fig:approximate_reuse_cases} illustrates this distinction: the shared interface regions marked in blue are selected for reuse, while the table region marked in orange is selected for recomputation.

\section{System Design}
\subsection{Overview}
\definecolor{qg}{RGB}{0,122,94}

\begin{figure}[t]
\centering
\includegraphics[width=\linewidth]{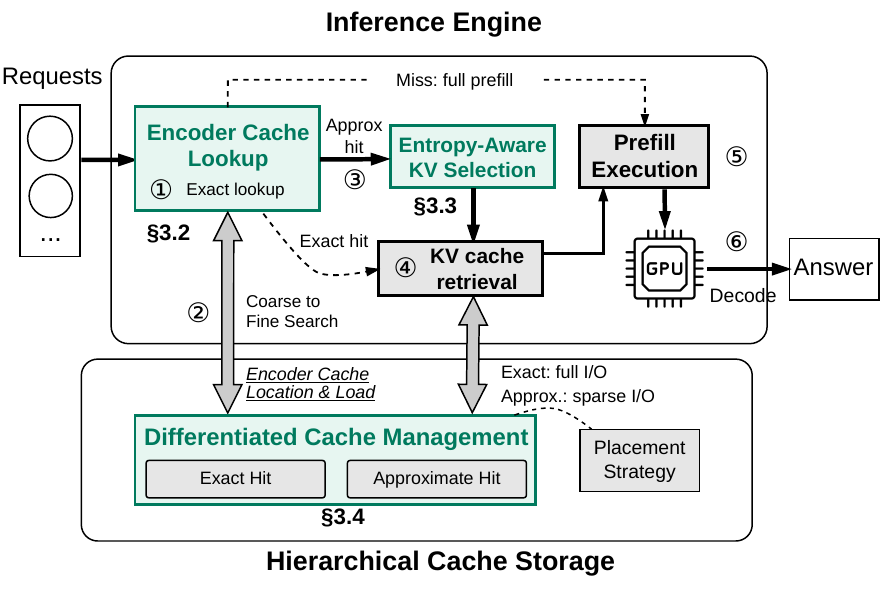}
\caption{Overview of \system.}
\label{fig:pipeline}
\end{figure}

\system\ extends visual KV cache reuse in VLM serving systems beyond exact matching to support approximate matches. As illustrated in~\autoref{fig:pipeline}, \system\ comprises three main components (\textcolor{qg}{\textbf{green blocks}}): \emph{Two-Stage Encoder Cache Lookup}, \emph{Entropy-Aware KV Selection} and \emph{Differentiated Cache Management}. We first provide an overview of each component and detail its design in the following subsections.

\noindent\textbf{Two-Stage Encoder Cache Lookup} identifies any ``available'' cached entries whose visual KV cache can be reused for a given visual input. Here, ``available'' refers either to exact-matched images or to visually approximate cached aliases retrieved via two-stage semantic approximate matching ($\S\ref{3.2}$) after an exact-match miss.

\noindent\textbf{Entropy-Aware KV Selection} determines the associated KV cache entries that can be reused and those that must be recomputed for a matched encoder-cache alias. Entries requiring recomputation are selected by probing attention distributions in warmup layers, whereas reusable entries are selected based on entropy theory  ($\S\ref{3.3}$).

\noindent\textbf{Differentiated Cache Management} serves exact and approximate hits with distinct I/O ($\S\ref{sec:sparse_io}$) and cache policies ($\S\ref{sec:cache_retention}$) across the GPU, CPU, and storage tiers, as the two hit types differ in temporal locality and data volume. In addition, a unified storage backend is implemented that jointly manages L2 memory and L3 storage for KV offloading, thereby supporting a substantially larger KV cache footprint ($\S\ref{sec:unified_backend}$).

\noindent\textbf{Workflow}. Based on \system\, the general workflow of VLM inference is as follows: Given an incoming request, such as a text prompt with an associated image,
\noindent\circled{1} Each incoming multimodal request first performs an exact image lookup using the input image hash. 
\noindent\circled{2} Once the exact match misses, the encoder-cache lookup performs a coarse-to-fine search, with the differentiated cache manager locating and loading cached encoder features for comparison. An exact hit bypasses this search and proceeds directly to the KV cache retrieval. \noindent\circled{3} If an approximate match is found, the entropy-aware reuse module compares patches in the input image with the corresponding patches in the approximate image, treating patches with low entropy as low-risk candidates for KV reuse. For the remaining patches, it uses warmup attention to select high-attention patches for recomputation. 
\noindent\circled{4} The differentiated cache manager then handles requests according to their access characteristics: exact hits retrieve the complete KV cache and receive higher retention priority, while approximate hits retrieve only selected patches through sparse I/O. It further combines offline workload priors with online reuse observations to determine which preset entries are retained in the reserved CPU victim cache. \noindent\circled{5} Requests without a valid match instead follow the normal prefill path. After placing the required KV cache in the GPU cache, the SGLang engine skips visual prefill for exact hits, combines cached KV reuse with selective patch recomputation for approximate hits, and performs full visual prefill and cache insertion for misses. 
\noindent\circled{6} Subsequently, all types of requests enter the normal decoding path, which generates and returns the final answers.

\subsection{Two-Stage Encoder Cache Lookup}\label{3.2}
When a query arrives and its associated visual input is being encoded, the first thing \system\ does is check if any historical requests can be reused. A na\"{i}ve lookup that compares the query's patch embeddings against every cached entry is impractical: an encoder cache typically contains thousands of tokens, and full-granularity comparison requires loading the complete encoder cache and computing token-level similarities for all candidates, all on the prefill critical path. \system\ therefore searches for the approximate match with a \emph{coarse-to-fine lookup} to identify an entry whose visual KV cache can be reused. It first checks the hash key for exact matches; upon an exact miss, \system\ uses the current encoder output to search for approximate matches. 

\noindent\textbf{Coarse-grained Searching.} Patch-level comparison requires the query and a cached image to share the same token-grid geometry, so \system\ searches the group of cached images whose visual token count and grid aspect ratio match the query. Within this group, \system\ compares compact regional descriptors. The patch grid for each image is divided into four regions in a 2×2 layout. Patch embeddings are individually normalized and averaged within each region to produce four regional descriptors. A candidate passes the coarse filter only if the cosine similarity of every corresponding descriptor is greater than the threshold $\tau$. Passing candidates are ranked by their mean regional similarities, and at most k are retained (k=5 in our configuration). This stage filters out unsuitable candidates with only four descriptor comparisons per image, without accessing their full encoder embeddings.

\noindent\textbf{Fine-grained Ranking.} For each remaining candidate, \system\ retrieves the full encoder embeddings and computes the cosine similarity between the patch embeddings at each corresponding token position. The fine-ranking score is computed as the average cosine similarity over all visual tokens, and the candidate with the highest score is selected as the approximate image. A search without a usable candidate is a cache miss. A successful lookup provides a matched image for the entropy-aware algorithm in $\S\ref{3.3}$, which determines the patch-level reuse and recomputation strategy.

\subsection{Entropy-Aware KV Cache Selection}\label{3.3}
\begin{figure}[t]
\centering
\includegraphics[width=\linewidth]{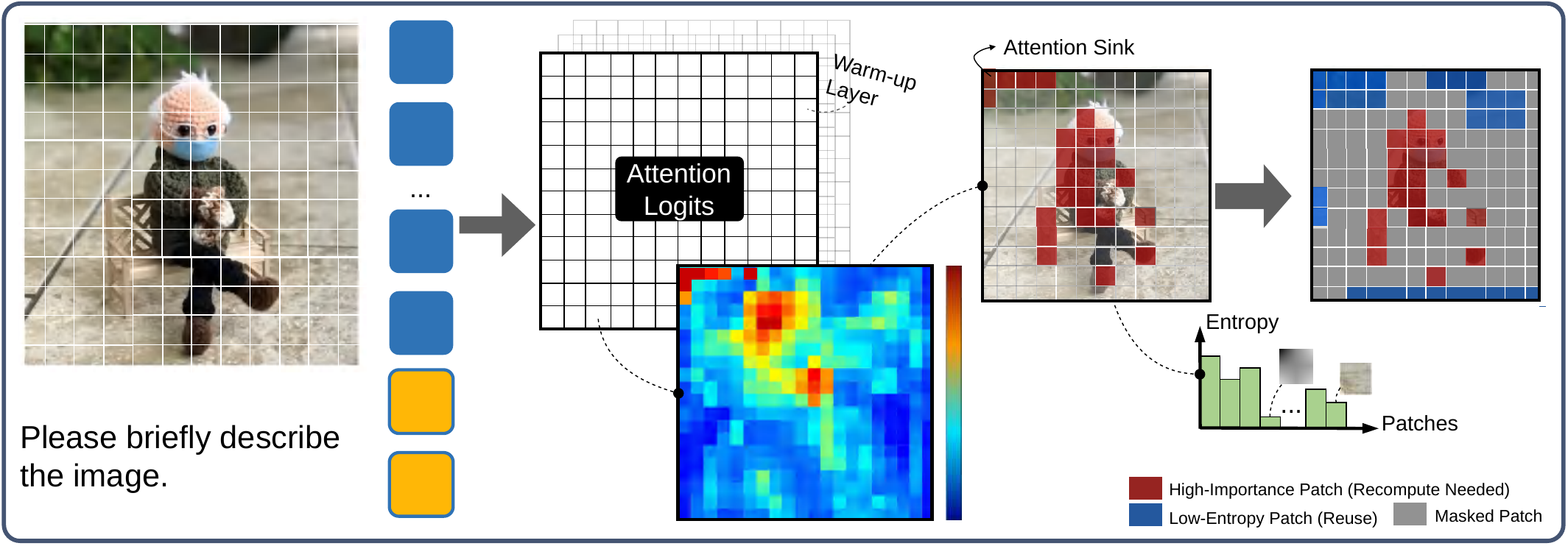}
\caption{Entropy-aware KV-cache selection process.}
\label{fig:entropy}
\end{figure}

Once the two-stage lookup identifies an approximate match, \system\ decides which visual patches to recompute, reuse, or mask in later transformer layers. As shown in~\autoref{fig:entropy}, it first identifies corresponding patches in the input and cached images as reuse candidates, then uses warmup attention to select patches for recomputation, and finally assigns the remaining patches to reuse or masking.

\subsubsection{Reuse Candidate Identification} \system\ leverages patch entropy to estimate reuse risk without considering relevance to the current prompt. Approximate images commonly share blank margins, uniform backgrounds, and other regions with little visual variation, whose KV cache is less risky to retain than patches containing high information density. For patch \(i\), let \(P_i\) denote its pixels and \(x_{p,c}\) the value of pixel \(p\) in channel \(c\). \system\ first measures the maximum channel-wise pixel range,
\[
D_i(x)=\max_c\left(\max_{p\in P_i}x_{p,c}-\min_{p\in P_i}x_{p,c}\right).
\]
Here, \(\delta\) denotes a predefined threshold for pixel-level variation. A patch is considered \emph{uniform} when \(D_i(x)\leq\delta\), indicating that its pixel values have limited variation across all color channels. However, this pixel-range selection only filters out visually rich regions. For each uniform patch, \system\ further computes its \emph{grayscale Shannon entropy}~\cite{shannon1948mathematical} to distinguish among the remaining uniform patches,
\[
H_i(x)=-\sum_{v=0}^{255}p_{i,v}(x)\log_2 p_{i,v}(x),
\]
where \(p_{i,v}(x)\) is the fraction of pixels in patch \(i\) with grayscale value \(v\). Based on these entropy values, \system\ stores the entropy scores and a bounded list of the lowest-entropy uniform patch indices for each image. For an approximate hit, let \(\mathcal{L}(a)\) denote the stored low-information patch indices of the matched image \(a\). A patch enters the paired candidate set only if its index belongs to \(\mathcal{L}(a)\) and the corresponding patch in the input image \(x\) is also uniform. This paired condition prevents stale reuse when a location that is blank in the cached image contains newly added text, a number, or an object in the input image. The paired candidate set is
\[
\mathcal{C}_0=\{i\in\mathcal{L}(a)\mid D_i(x)\leq\delta\}.
\]
\system\ assigns each candidate a combined entropy score \(E_i(x,a)=H_i(x)+H_i(a)\). A low value indicates that the corresponding patch contains little visual variation in both images and is therefore a conservative candidate for KV reuse.

\begin{figure}[t]
\centering
\begin{minipage}{0.49\textwidth}
\centering
\begin{subfigure}{0.49\linewidth}
    \includegraphics[width=\linewidth]{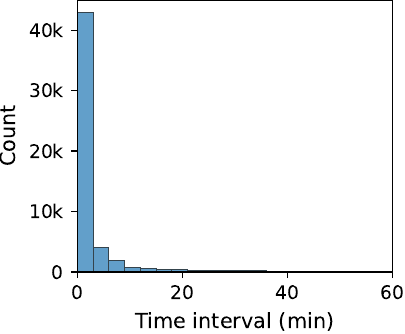}
    \caption{Exact-hit intervals}
    \label{fig:exact_hit_interval_distribution}
\end{subfigure}
\hfill
\begin{subfigure}{0.49\linewidth}
    \includegraphics[width=\linewidth]{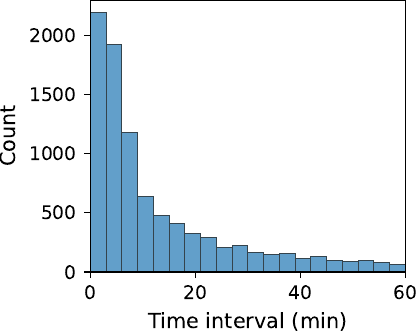}
    \caption{Approximate-hit intervals}
    \label{fig:approx_hit_interval_distribution}
\end{subfigure}
\end{minipage}
\vspace{-3mm}
\caption{Distributions of cache-hit time intervals. Exact-hit intervals are predominantly short, whereas approximate-hit intervals exhibit a broader, heavier-tailed distribution.}
\label{fig:cache_hit_interval_distribution}
\end{figure}

\subsubsection{Recompute Token Selection}
Entropy alone cannot determine which patches matter to the current request. \system\ executes the first \(W\) transformer layers normally and measures the attention received by each visual patch at the last warmup layer. Let \(A_{j,i}^{(h)}\) denote the attention from causal query position \(j\) to visual patch \(i\) at head \(h\), and let \(\mathcal{Q}_i\) contain the query positions that can attend to patch \(i\). \system\ computes the normalized importance score
\[
I_i=\frac{1}{H|\mathcal{Q}_i|}\sum_{h=1}^{H}\sum_{j\in\mathcal{Q}_i}A_{j,i}^{(h)},
\]
where \(H\) is the number of attention heads. Normalizing by \(|\mathcal{Q}_i|\) removes the positional bias caused by earlier visual patches being visible to more subsequent positions. The resulting score reflects the attention assigned to each patch under the current request and guides computation in the remaining layers.

\system\ combines the two signals by constructing the recompute set before the reuse set. Given a recomputation ratio \(\rho\), it allocates \(k_{\mathrm r}=\lceil\rho N\rceil\) patches and first selects the highest-attention patches from the non-uniform set \(\mathcal{V}(x)=\{i\mid D_i(x)>\delta\}\):
\[
\mathcal{R}=\operatorname{TopK}_{i\in\mathcal{V}(x)}(I_i,k_{\mathrm r}).
\]
If \(|\mathcal{V}(x)|<k_{\mathrm r}\), the remaining budget is filled using the highest-attention patches outside \(\mathcal{V}(x)\). This prevents uniform backgrounds from consuming the recomputation budget when they receive spuriously high attention, thus preventing the visual token sink problem.

\subsubsection{Reuse and Mask Assignment}
Given a reuse ratio \(\beta\), \system\ then removes \(\mathcal{R}\) from the paired candidates and selects the patches with the lowest combined entropy:
\[
\begin{aligned}
\mathcal{C}&=\mathcal{C}_0\setminus\mathcal{R},\\
\mathcal{B}&=\operatorname{BottomK}_{i\in\mathcal{C}}
\left(E_i(x,a),\min\{\lceil\beta N\rceil,|\mathcal{C}|\}\right).
\end{aligned}
\]
In the remaining transformer layers, \system\ recomputes patches in \(\mathcal{R}\), reuses the KV cache of patches in \(\mathcal{B}\), and masks all other visual patches by a sparse attention mask. Recompute takes priority over reuse, allowing a low-entropy patch to be refreshed whenever the current request assigns it sufficient attention.

\begin{figure}[t]
\centering
\includegraphics[width=\linewidth]{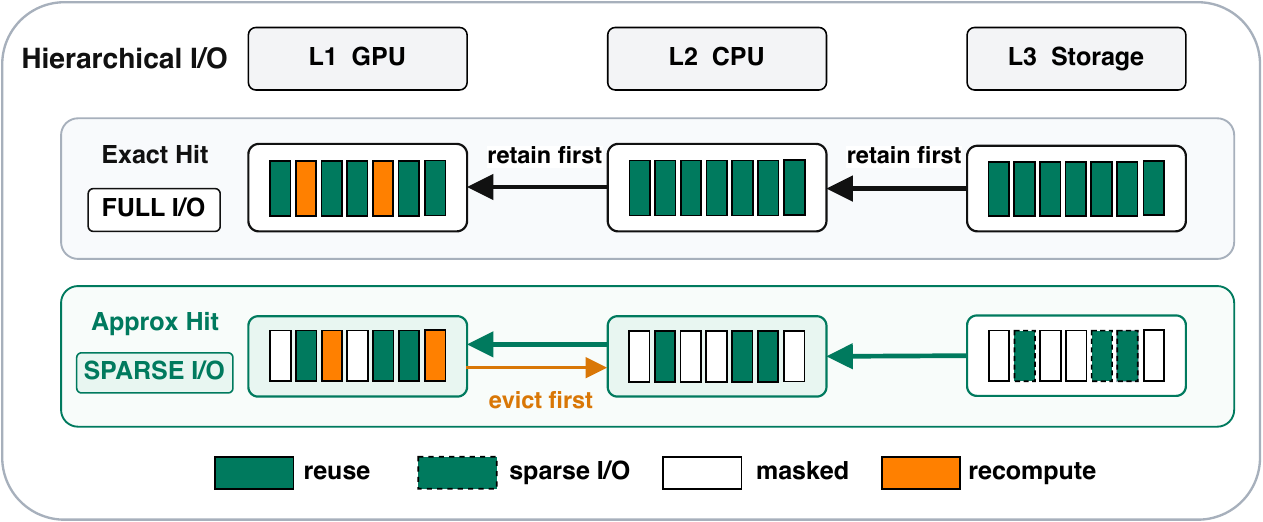}
\caption{Differentiated hierarchical I/O in \system. Exact hits use full I/O, while approximate hits use sparse I/O to retrieve only the selected patches.}
\label{fig:differentiated_hierarchical_io}
\end{figure}

\subsection{Differentiated Hierarchical Cache Management}\label{3.4}
Patch selection determines which KV cache an approximate hit needs, while cache management further determines how to retrieve and retain them. In real-world VLM serving workloads, exact and approximate hits exhibit substantially different temporal locality. \autoref{fig:cache_hit_interval_distribution} shows that approximately 82\% of exact hits occur within the first 3-minute interval. In contrast, only about 46\% of approximate hits occur within 6 minutes, while more than 20\% occur after 21 minutes. Exact hits therefore benefit from retention in the upper cache tiers, whereas approximate reuse requires a broader capacity across CPU memory and SSDs. The two cache-hit types also move different amounts of data: an exact hit consumes the complete visual KV cache, while an approximate hit consumes only patches selected by the entropy-aware reuse algorithm. Applying the same I/O and eviction policy to both would pollute the upper cache tiers and allow approximate reuse to displace entries with stronger exact-reuse locality. \system\ therefore differentiates full I/O, sparse I/O, and retention decisions across its GPU, CPU, and storage tiers.

\subsubsection{Patch-Level Sparse I/O}
\label{sec:sparse_io}
As shown in Figure~\ref{fig:differentiated_hierarchical_io}, exact hits use full I/O to retrieve the complete visual KV cache into the current request's GPU KV slots, loading SSD-resident entries into the CPU cache first. Approximate hits compute the first \(W\) warmup layers normally and use sparse I/O to retrieve only the subsequent-layer KV cache for patches in the reuse set \(\mathcal{B}\). The respective transfer volumes are
\[
\begin{aligned}
V_{\mathrm{full}} &= 2LNdb,\\
V_{\mathrm{sparse}} &= 2(L-W)|\mathcal{B}|db,
\end{aligned}
\]
where \(L\) is the total number of transformer layers, \(N\) the number of visual patches, \(d\) the hidden dimension of KV vector, and \(b\) the bytes per element.

The cache hierarchy uses a patch-first layout in CPU memory and SSD storage to directly locate each patch's complete KV cache. For sparse I/O, the CPU tier transfers only the selected patches, while the SSD tier reads only the corresponding byte ranges. The complete alias entry remains in its original tier instead of being promoted because of partial access, reducing both transfer traffic and cache pollution. Figure~\ref{fig:cum_io_sparse} shows the cumulative I/O of sparse and full cache retrieval under the real-world request trace.

\begin{figure}[t]
\centering
\includegraphics[width=\linewidth]{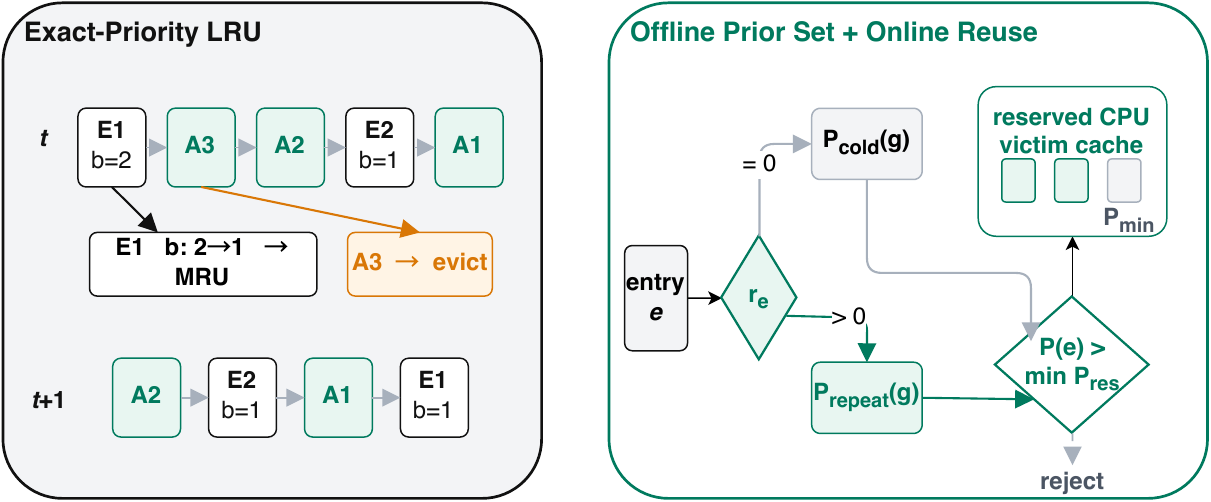}
\caption{Differentiated cache retention in \system: exact-priority LRU (left) and CPU victim cache retention using offline priors and online reuse observations (right). }
\label{fig:differentiated_cache_retention}
\end{figure}

\subsubsection{Cache Retention Policies}
\label{sec:cache_retention}
Sparse I/O reduces the KV cache transfer latency, but limited cache capacity still requires careful retention decisions~\cite{jin2026ragcache}. Standard least-recently used (LRU) eviction exploits temporal locality by ordering entries according to their most recent access. However, treating exact and approximate hits equally ignores the shorter reuse intervals of exact hits shown in~\autoref{fig:cache_hit_interval_distribution}.
\system\ therefore uses \emph{exact-priority LRU}, as illustrated in~\autoref{fig:differentiated_cache_retention} (left). Each exact-hit entry receives a bounded protection budget: when eviction reaches a protected entry, its budget is decremented and the entry is moved to the most-recently used position. Approximate-only entries are evicted first. This bounded budget prevents stale exact entries from remaining resident indefinitely, while reducing displacement of KV caches with stronger full-reuse locality.

\begin{figure}[t]
    \centering
    \begin{subfigure}{0.49\columnwidth}
        \centering
        \includegraphics[width=\linewidth]{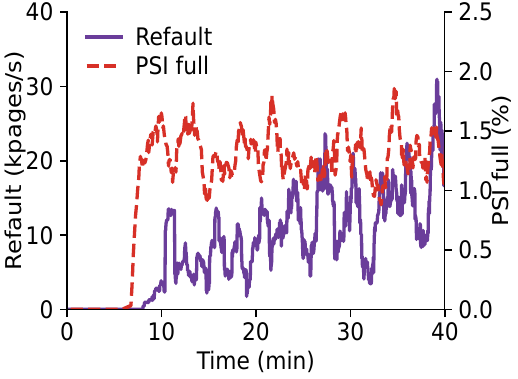}
        \caption{Mooncake memory pressure}
        \label{fig:mooncake_pressure}
    \end{subfigure}
    \hfill
    \begin{subfigure}{0.49\columnwidth}
        \centering
        \includegraphics[width=\linewidth]{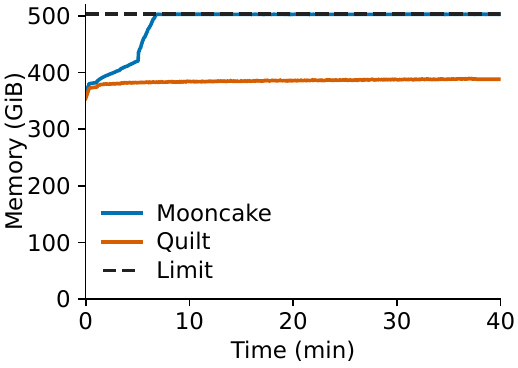}
        \caption{Total memory usage}
        \label{fig:total_memory}
    \end{subfigure}
    \vspace{\baselineskip}  
    \begin{subfigure}{0.49\columnwidth}
        \centering
        \includegraphics[width=\linewidth]{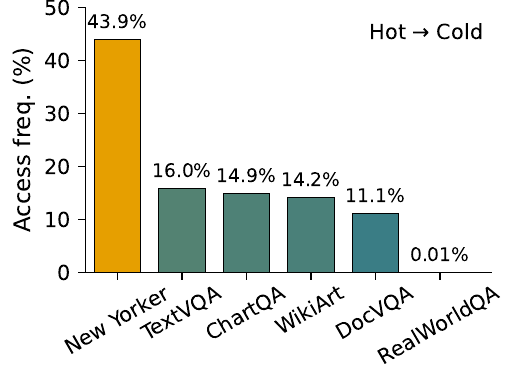}
        \vspace{-6mm}
        \caption{Preset label access frequency}
        \label{fig:preset_label_access_frequency}
    \end{subfigure}
    \hfill
    \begin{subfigure}{0.49\columnwidth}
        \centering
        \includegraphics[width=\linewidth]{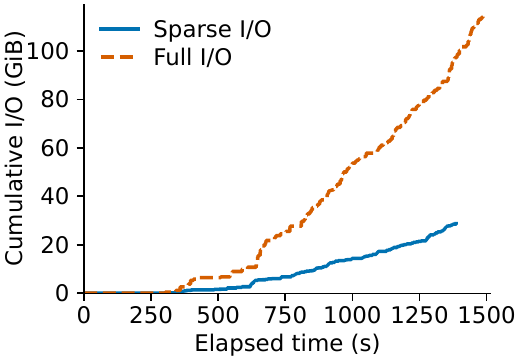}
        \vspace{-6mm}
        \caption{Cumulative sparse I/O}
        \label{fig:cum_io_sparse}
    \end{subfigure}
    \vspace{-6mm}
    \caption{Memory pressure, preset image access frequency, and I/O volume in VLM workloads.}
    \label{fig:combined}
\end{figure}

Exact-priority LRU favors entries with short-term exact reuse, while approximate reuse can benefit from retaining entries over longer intervals. In real-world workloads, preset images are provided in advance for users to select, and their accesses are scattered over time with weak temporal locality. \autoref{fig:preset_label_access_frequency} shows that preset accesses are strongly skewed across the six categories, with access frequencies ranging from 43.9\% to 0.01\%. To extend approximate-match coverage without polluting the general CPU cache, \system\ reserves a CPU victim cache for preset entries and manages it using offline workload priors and online reuse observations. Offline traces estimate two reuse priors for each category and vision-token bucket: a cold prior for entries without observed reuse and a repeat prior for entries that have already been reused\footnote{The prior training trace is completely separated from the evaluation trace.}. The priors are fixed before the evaluation and do not use future access records from the evaluation trace. Each online request includes its workload class and vision-token bucket, which determine \(g(e)\), while \(r_e\) is updated only from reuse observations available before the current retention decision. As shown in~\autoref{fig:differentiated_cache_retention} (right), the online reuse count selects the applicable prior. For an entry \(e\), the retention priority is
\[
P(e)=
\begin{cases}
P_{\mathrm{cold}}(g(e)), & r_e=0,\\
P_{\mathrm{repeat}}(g(e)), & r_e>0,
\end{cases}
\]
where \(g(e)\) is the workload bucket of the entry and \(r_e\) is its observed online reuse count. A new entry is admitted when the CPU victim cache has enough free space. Otherwise, it replaces the lowest-priority resident entries needed to make room only if its priority exceeds each of theirs. The cold prior addresses the absence of online history, while the repeat prior incorporates evidence that an entry has been useful in the running workload. Together, sparse I/O bounds the cost of approximate reuse, exact-aware protection preserves entries with strong full-reuse locality, and the reserved victim cache expands long-tail approximate coverage.

\subsubsection{Unified Memory and Storage Backend}
\label{sec:unified_backend}
\begin{figure}[t]
\centering
\includegraphics[width=\linewidth]{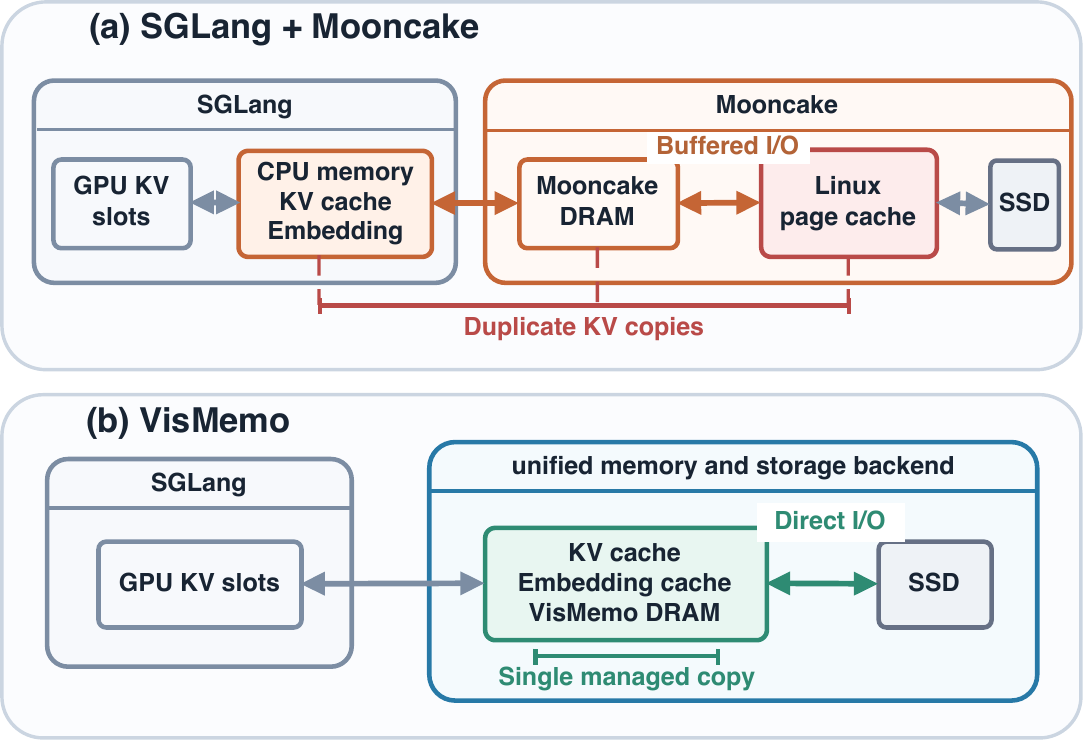}
\vspace{-6mm}
\caption{Differences in memory access paths between Mooncake and \system\ integrated with SGLang.}
\label{fig:unified_memory_access_path}
\end{figure}

We consider a common deployment in which SGLang serves as the inference engine, and Mooncake provides the storage backend~\cite{zheng2024sglang,qin2025mooncake,chen2026generativetransmission}. In this case, the resulting cache hierarchy consists of GPU HBM and privately-owned CPU memory managed by SGLang, as well as the remaining CPU memory and SSDs managed by Mooncake. This hierarchy introduces three sources of overhead for VLM workloads. First, as shown in~\autoref{fig:unified_memory_access_path}, the Mooncake SSD backend uses \emph{buffered I/O}, causing written data to enter the Linux page cache~\cite{mooncake2026repository}. As the explicitly managed DRAM cache and the page cache grow simultaneously, the total memory usage gradually approaches the limit. As shown in~\autoref{fig:mooncake_pressure}, frequent page faults and reclamation can stall inference and increase long-tail prefill latency. Second, after KV caches are fetched from Mooncake memory to SGLang CPU memory, both levels retain identical host-to-host copies. This duplication reduces the effective cache capacity and introduces additional data movement. Finally, SGLang does not implement a dedicated memory pool for the encoder cache~\cite{sglang2026repository}. Thus, encoder caches reside mainly in the GPU and Mooncake. Once a miss occurs, the data must first be read into a pinned staging buffer and then copied to HBM, introducing an additional memory copy and staging-slot wait time.

To eliminate these redundancies, \system\ designs a unified and explicitly managed memory access pattern. The KV cache and the embedding cache share a unified memory pool in \system, replacing duplicate copies previously maintained in SGLang memory and Mooncake DRAM. The storage backend uses variable-size allocation to accommodate objects generated by different numbers of visual tokens. It also centrally manages indexing, placement, eviction, and SSD-to-DRAM promotion, preventing the configured DRAM budget from being consumed by redundant copies across different software layers.

\input{tables/accuracy}

For the KV cache, \system\ registers the backend DRAM as CUDA pinned memory and stores KV cache in a patch-first layout. For writes, a customized kernel transfers KV cache directly from GPU memory to its final DRAM allocation. For reads, the backend returns an object view from which the restoration kernel can either reconstruct the complete KV cache or selectively restore the KV cache of later layers for patches chosen for approximate reuse. For contiguous embedding caches, \system\ uses the \emph{CUDA copy engine} to asynchronously transfer data between the unified memory pool and GPU memory, eliminating the intermediate staging copy required by Mooncake. In the SSD layer, \system\ uses direct I/O to bypass the page cache managed by the operating system. It manages SSD data through asynchronous batched writes, parallel I/O across multiple shards, and explicit DRAM admission control. \system\ explicitly determines whether a hot object remains in DRAM or is promoted from SSD, rather than relying on uncontrolled kernel page residency. This design reduces redundant caching and unnecessary data movement, while isolating inference from OS-driven page reclamation, thereby ensuring predictable memory usage and tail latency.

\section{Evaluation}
\subsection{Evaluation Setup}

\noindent\textbf{Software and Hardware}. All experiments are conducted on a server equipped with an Intel Xeon Platinum processor with 48 cores, 500~GB of DRAM, and two NVIDIA H100 GPUs, each with 80~GB of HBM, connected via NVLink with a bidirectional bandwidth of 400~GB/s. The server runs Ubuntu 24.04 with CUDA 12.8, PyTorch 2.9.1. \system\ is implemented on SGLang~\cite{zheng2024sglang} and uses FlashInfer~\cite{ye2025flashinfer} as the attention backend. We run the backend VLMs with tensor parallelism of 2. For KV cache capacity in each TP rank, we allocate 10~GB of GPU memory to L1 and 100~GiB of CPU memory to L2. L3 is configured with 4~TB of SSDs. For the memory pressure experiment ($\S\ref{4.4}$), we impose a 500~GB CPU memory limit and configure \system\ with 4~TB of SSD capacity.

\noindent\textbf{Models and Workloads}. We employ \emph{Qwen3-VL-32B}~\cite{bai2025qwen3vl} as our primary model and additionally evaluate \emph{Qwen3-VL-8B}~\cite{bai2025qwen3vl} and \emph{GLM-4.6V-Flash}~\cite{vteam2025glm} in the system efficiency evaluation ($\S\ref{4.3}$). We use the real-world VLM conversational workload \emph{VisionArena}~\cite{chou2025visionarena}, which preserves the original request order and temporal locality, along with four public datasets, \emph{InfographicVQA}~\cite{mathew2022infographicvqa}, \emph{DocVQA}~\cite{mathew2021docvqa}, \emph{MMBench}~\cite{liu2024mmbench}, and \emph{SEED-Bench}~\cite{li2023seedbench}, which cover open-ended visual queries, document understanding, and multimodal reasoning.

\noindent\textbf{Baselines}. We use \emph{VLCache}~\cite{qin2025vlcache} and \emph{CacheBlend}~\cite{yao2025cacheblend} as baselines to evaluate output quality, while evaluating \system\ against \emph{Mooncake}~\cite{qin2025mooncake} for system efficiency. All comparisons use the same request order, prompts, model configuration, cache capacities, and output length limit.

\noindent\textbf{Metrics}. We evaluate both quality and efficiency. For quality, we report \emph{average normalized levenshtein similarity (ANLS)}~\cite{biten2019scene}, which measures the textual similarity between a model’s prediction and the ground-truth for VisionArena, InfographicVQA, and DocVQA, and \emph{accuracy} for others. For efficiency, we compare with Mooncake in terms of \emph{average prefill latency} for exact and approximate hits. We further report \emph{cache hit distributions} across tiers and analyze latency savings from individual system components.

\noindent\textbf{Implementation Details}. We retain up to five candidates from cache lookup with cosine similarity above $\tau=0.8$. For approximate hits, we use $W=3$ warmup layers with a recomputation ratio of $\rho=0.3$ and a reuse ratio of $\beta=0.1$.
  
\begin{figure*}[t]
    \centering
    \captionsetup[subfigure]{labelformat=empty}
    \begin{subfigure}[t]{0.29\textwidth}
        \caption{Qwen3-VL-8B}
        \label{fig:prefill_latency_qwen3_8b}
        \includegraphics[width=\linewidth]{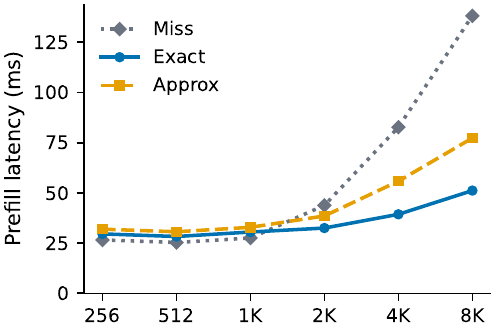}
    \end{subfigure}
    \hfill
    \begin{subfigure}[t]{0.29\textwidth}
        \caption{Qwen3-VL-32B}
        \label{fig:prefill_latency_qwen3_32b}
        \includegraphics[width=\linewidth]{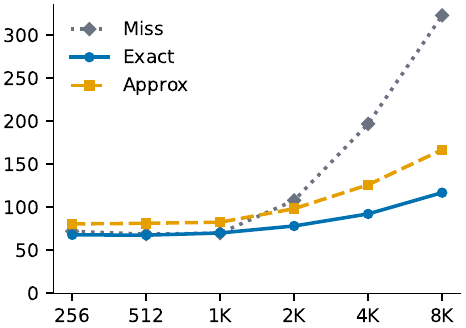}
    \end{subfigure}
    \hfill
    \begin{subfigure}[t]{0.29\textwidth}
        \caption{GLM-4.6V-Flash}
        \label{fig:prefill_latency_glm46v}
        \includegraphics[width=\linewidth]{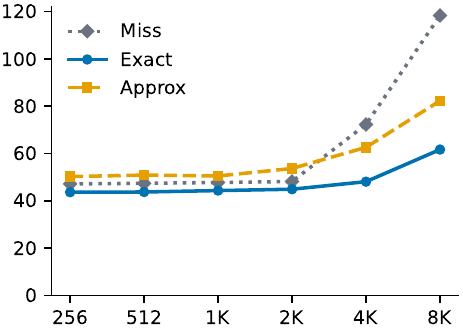}
    \end{subfigure}

    \begin{subfigure}[t]{0.29\textwidth}
        \includegraphics[width=\linewidth]{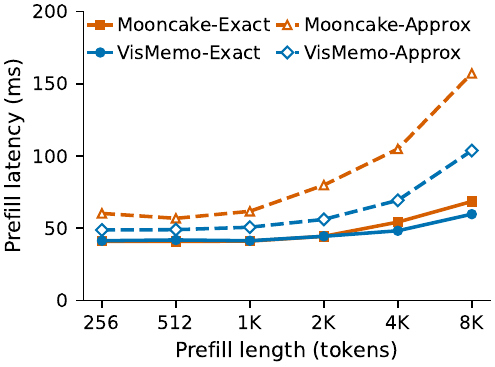}
        \phantomsubcaption\label{fig:prefill_latency_mooncake_vs_quilt_qwen3_8b}
    \end{subfigure}
    \hfill
    \begin{subfigure}[t]{0.29\textwidth}
        \includegraphics[width=\linewidth]{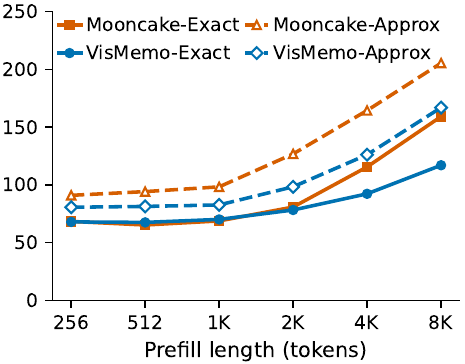}
        \phantomsubcaption\label{fig:prefill_latency_mooncake_vs_quilt_qwen3_32b}
    \end{subfigure}
    \hfill
    \begin{subfigure}[t]{0.29\textwidth}
        \includegraphics[width=\linewidth]{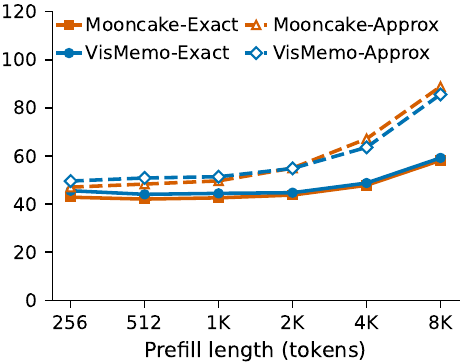}
        \phantomsubcaption\label{fig:prefill_latency_mooncake_vs_quilt_glm46v}
    \end{subfigure}
    \vspace{-6mm}
    \caption{Prefill latency across varying input lengths (top row) and comparison of Mooncake (bottom row) on VisionArena.}
    \label{fig:prefill_latency_combined}
\end{figure*}

\subsection{Overall Quality Improvement}
Table~\ref{tab:accuracy} reports the output quality of Qwen3-VL-8B, with \system\ achieving the highest overall scores on every dataset. Compared with the strongest baseline on each dataset, the overall gains are 0.19 points on VisionArena, 1.14 points on InfographicVQA, 11.64 points on DocVQA, 0.23 points on MMBench, and 1.15 points on SEED-Bench. Notably, the largest gain is on DocVQA, where \system\ reaches an ANLS of 64.30\textemdash 22.1\% higher than VLCache and 45.2\% higher than CacheBlend. Across the five datasets, \system\ averages 70.81 overall and 64.21 on approximate hits, surpassing the best baseline averages by 4.77 and 15.17 points, respectively.

The approximate-hit subset directly measures the quality of reusing a non-identical image. \system\ obtains the highest approximate-hit score in the five datasets. Compared with the strongest baseline for each dataset, it improves the accuracy of approximate hits by 16.14 points in VisionArena, 19.72 points in InfographicVQA, and 17.27 points in DocVQA. The approximate-hit accuracy improves by 1.36 points in MMBench and 5.33 points in SEED-Bench. For exact hits, \system\ remains within 0.68 points of the best exact-hit baseline on all datasets\footnote{Minor exact-hit fluctuations are expected because some exact hits are classified as approximate hits by \system.}. These results show improved approximate-hit quality and comparable exact-hit quality of \system.

\subsection{System Efficiency}\label{4.3}
\autoref{fig:prefill_latency_combined} shows that the benefit of visual KV cache reuse increases with visual input length under Qwen3-VL-8B, Qwen3-VL-32B, and GLM-4.6V-Flash. At 8K input tokens, cache miss requests take 138.2, 323.0, and 118.3~ms of average prefill latency on the three models, respectively. Exact hits reduce these latencies to 51.1, 117.0, and 61.7~ms, corresponding to 2.7$\times$, 2.8$\times$, and 1.9$\times$ speedups. The approximate hits take 76.5, 166.9, and 82.3~ms, reducing the miss latency by 44.6\%, 48.3\%, and 30.5\%, respectively. At 2K tokens, approximate reuse reduces latency from 43.8 to 38.5~ms on Qwen3-VL-8B and from 108.0 to 98.3~ms on Qwen3-VL-32B. On GLM-4.6V-Flash, the approximate-hit latency is 53.7~ms compared to 48.2~ms for a miss at 2K tokens, but it falls below the miss latency at 4K tokens, reducing the latency from 72.3 to 62.6~ms. These results show that the crossover point depends on the model, while exact hits consistently achieve the lowest latency because they avoid warmup and recomputation.

Combined with Table~\ref{tab:accuracy}, these results show that \system\ converts the additional approximate hits into prefill savings while maintaining the higher output quality for requests served through approximate reuse. The benefit becomes pronounced once the avoided prefill computation dominates the fixed costs of approximate lookup, warmup attention, and selective recomputation.

\autoref{fig:prefill_latency_combined} compares Mooncake with \system's unified memory and storage backend across the three models. On Qwen3-VL-8B, the exact-hit latencies are nearly identical through 2K tokens. At 4K and 8K tokens, \system\ reduces exact-hit latency from 54.2 to 48.2~ms and from 68.5 to 59.7~ms, corresponding to reductions of 11.1\% and 12.8\%. The improvement is larger for approximate hits: at 2K, 4K, and 8K tokens, latency decreases from 79.8 to 56.1~ms, from 104.9 to 69.3~ms, and from 157.2 to 103.7~ms, respectively.

On Qwen3-VL-32B, exact-hit latencies remain within 3.5\% through 2K tokens, but the gap widens as the visual KV cache grows. At 4K and 8K tokens, \system\ reduces latency from 115.6 to 92.3~ms and from 159.2 to 117.0~ms, corresponding to reductions of 20.2\% and 26.5\%. For approximate hits, \system\ is faster at every input length. At 2K, 4K, and 8K tokens, it reduces latency by 22.4\%, 23.3\%, and 18.7\%, respectively.

On GLM-4.6V-Flash, the two backends have a similar exact-hit latency. Approximate-hit latency is also similar for short inputs and begins to favor \system\ at longer lengths. At 4K and 8K tokens, \system\ reduces approximate-hit latency from 67.1 to 63.6~ms and from 88.9 to 85.6~ms, corresponding to reductions of 5.3\% and 3.7\%. Across the three models, the unified memory and storage backend provides the largest gains for the Qwen models while maintaining comparable performance on GLM-4.6V-Flash.

Exact and approximate hits exercise different parts of the design. Exact hits use full I/O to retrieve complete entries. Their long-input gains come from the unified memory pool and the restoration kernel, which eliminate duplicate copies. For approximate hits, the patch-first layout and the sparse I/O at patch granularity retrieve only the late-layer KV cache associated with the reuse set \(\mathcal{B}\). The complete alias entry remains in its original tier instead of being promoted after partial access. These results show that the differentiated full I/O and sparse-I/O paths reduce memory access overhead by offsetting the computational savings of visual KV cache reuse as the visual sequence grows.

\begin{figure}[t]
\centering
\includegraphics[width=\linewidth]{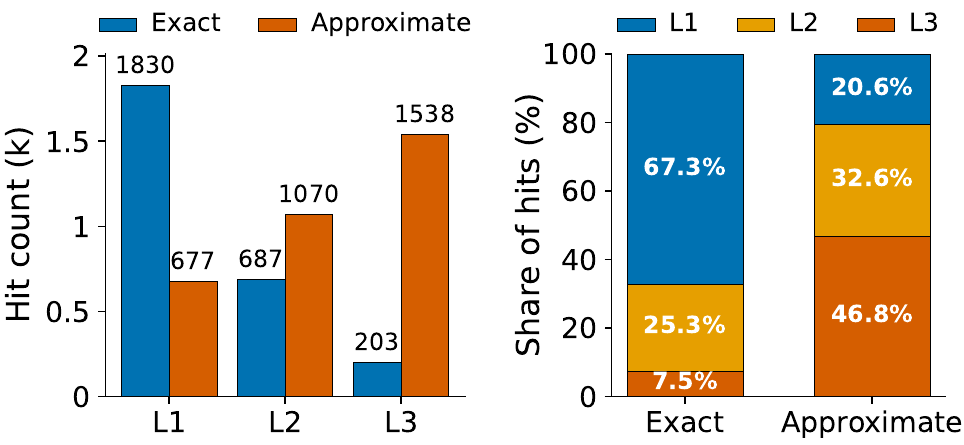}
\vspace{-6mm}
\caption{Cache-tier usage by hit type: cache hit counts (left) and tier-wise hit shares (right).}
\label{fig:cache_tier_hits}
\end{figure}

\subsection{Performance Evaluation}\label{4.4}
\noindent\textbf{Cache Behavior}.
In the VisionArena trace, approximate matching increases the overall cache hit rate from 27.23\% to 60.28\%, adding 33.05 percentage points beyond exact matching. \autoref{fig:cache_tier_hits} shows that the two types of hits have distinct distributions across the cache hierarchy. Here, L1 is the GPU memory tier, while L2 and L3 are the CPU memory and SSD-backend storage tiers managed by \system. The exact hits are concentrated in the upper tiers: 67.3\% are served by L1, and L1 and L2 together account for 92.5\%. The approximate hits depend more on the lower storage tier, with 79.4\% served by L2 or L3, and L3 alone accounting for 46.8\%. This difference is consistent with the short exact-hit intervals and the long-tailed approximate-hit intervals in~\autoref{fig:cache_hit_interval_distribution}. These access patterns support prioritizing exact entries for upper-tier retention while using the larger capacity of lower tiers to extend approximate reuse.

\begin{figure}[t]
\centering
\includegraphics[width=\linewidth]{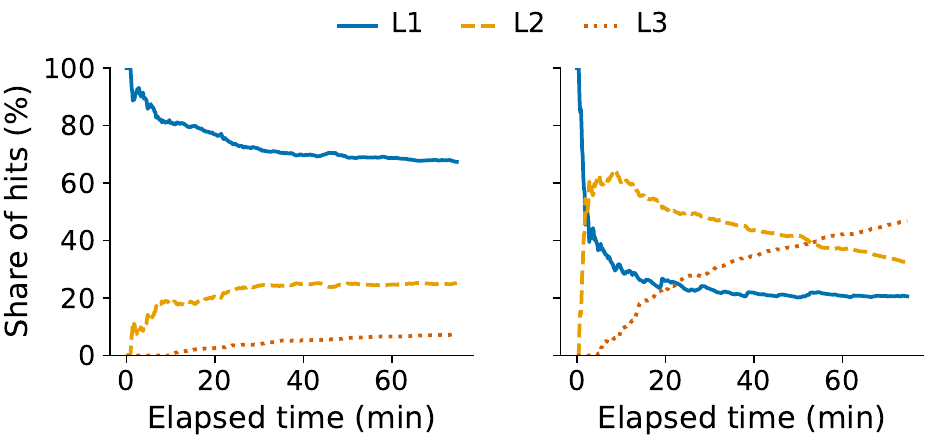}
\vspace{-7mm}
\caption{Cache-hit distributions over the VisionArena trace: exact hits (left) and approximate hits (right).}
\label{fig:hit_rate_over_time}
\end{figure}

The tier distribution evolves differently over the trace. As shown in~\autoref{fig:hit_rate_over_time}, the L1 share of exact hits remains dominant and converges to 67.3\%, while L2 and L3 stabilize at 25.3\% and 7.5\%. The approximate hits initially concentrate on L1 and L2, but the L3 share increases over time. L3 overtakes L2 after approximately 50 minutes and ends at 46.8\%, compared to 32.6\% for L2 and 20.6\% for L1. Approximate reuse exhibits weaker temporal locality, making matching entries less likely to remain in the upper tiers. The larger capacity of L3 allows it to retain more cached images, providing a broader candidate pool for approximate matching. This behavior is consistent with our insight in $\S\ref{3.3}$: exact reuse exhibits stronger temporal locality and benefits from upper-tier retention, while approximate reuse benefits from the larger pool of cached images retained in SSDs.

\begin{figure*}[t]
\centering
\begin{minipage}[t]{0.3\textwidth}
\centering
\vspace{0pt}
\includegraphics[width=\linewidth]{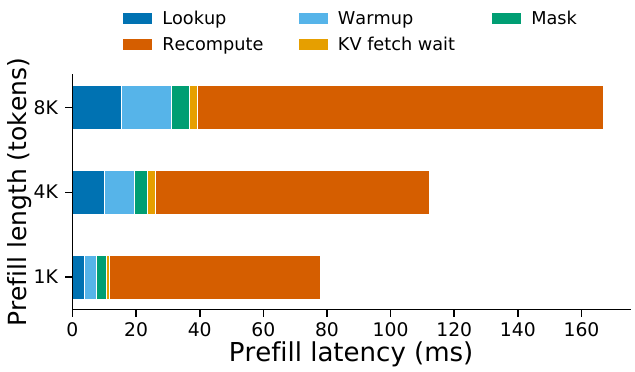}
\vspace{-6mm}
\caption{Approximate-hit prefill latency breakdown for Qwen3-VL-32B with 1K, 4K, and 8K-token inputs.}
\label{fig:performance_breakdown}
\end{minipage}
\hfill
\begin{minipage}[t]{0.3\textwidth}
\centering
\vspace{0pt}
\includegraphics[width=\linewidth]{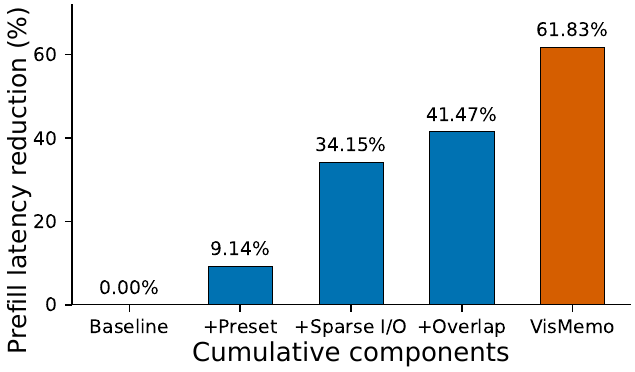}
\vspace{-6mm}
\caption{Cumulative approximate-hit prefill latency reduction for Qwen3-VL-32B relative to baseline.}
\label{fig:component_breakdown}
\end{minipage}
\hfill
\begin{minipage}[t]{0.3\textwidth}
\centering
\vspace{0pt}
\includegraphics[width=\linewidth]{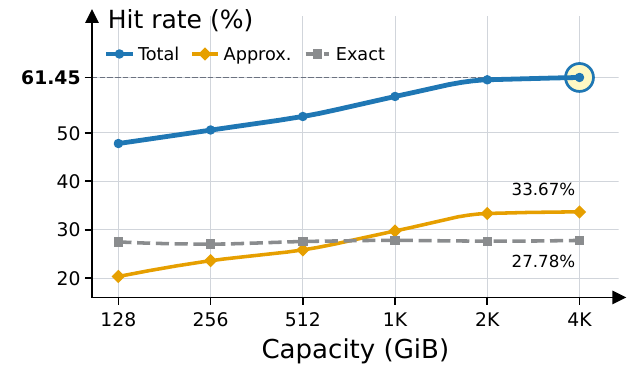}
\vspace{-6mm}
\caption{Cache hit rates with SSD capacity scales, where Total is the sum of exact and approximate hit rates.}
\label{fig:ssd_scale_hit_rate}
\end{minipage}
\end{figure*}

\noindent\textbf{Performance Breakdown}.
Figure~\ref{fig:performance_breakdown} represents the prefill latency breakdown of approximate hits for different visual sequence lengths. Selective recomputation dominates the total prefill latency. Its cost increases from 67~ms in the 1K sequence length to 128~ms in the 8K sequence length, while the total prefill latency increases from 78~ms to 167~ms.

Lookup measures the time spent finding a similar cached image for approximate reuse. Its latency increases with visual sequence length as approximate matching requires retrieving and comparing more visual tokens. Warmup executes the first three layers to collect attention scores for selective recomputation. Its latency also increases with visual input. Together, lookup and warmup account for about 10\% of total prefill latency at 1K tokens and 19\% at 8K tokens, accounting for most of the remaining prefill latency. Mask construction and KV fetch wait remain relatively small. KV retrieval starts asynchronously and overlaps with the warmup to hide KV cache transfer latency from lower cache tiers. KV fetch wait records only the remaining stall before the KV cache can be used. The small waiting component therefore does not represent the full KV retrieval latency.

\autoref{fig:component_breakdown} shows the cumulative benefits of system optimizations for approximate-hit prefill of Qwen3-VL-32B. The baseline runs the approximate KV-reuse algorithm on SGLang with Mooncake. We progressively enable preset retention, sparse I/O, and I/O--warmup overlap, and the optimized memory and storage access path used in full \system\. Preset retention prioritizes KV entries to retain in the reserved CPU cache using offline workload priors and online reuse observations, reducing approximate-hit prefill latency by 9.14\%. Sparse I/O increases the cumulative reduction to 34.15\% by reading only the KV cache selected for reuse, while I/O–warmup overlap raises it to 41.47\% by overlapping L3 cache retrieval with warmup computation. Finally, the optimized access path, including direct I/O and eliminating redundant memory copies, reduces prefill latency to 122.04~ms\textemdash a cumulative reduction of 61.83\%.

\noindent\textbf{SSD Capacity Scaling}.
\autoref{fig:ssd_scale_hit_rate} shows the cache hit rates as the SSD capacity scales from 128~GB to 4~TB. The exact hit rate remains almost constant around 27.8\% across all capacities: exact reuse exhibits strong temporal locality, and the exact-priority retention policy further keeps repeated entries in the upper tiers, making exact hits insensitive to SSD capacity. In contrast, the approximate hit rate increases from 20.33\% at 128~GB to 33.35\% at 2~TB, overtaking the exact hit rate at 1~TB. A larger SSD retains more cached images and enlarges the candidate pool for approximate matching, confirming that approximate reuse is primarily capacity-bound. Both curves saturate at 2~TB where the total hit rate reaches 61.45\%. Beyond this point, the SSD capacity is no longer the bottleneck, and doubling it to 4 TiB adds only 0.48 percentage points, as the remaining misses find no approximate match passing the similarity threshold.
This scaling behavior confirms the capacity-scaling trend predicted by the simulation in~\autoref{fig:hit_rate_vs_capacity}.

\begin{figure}[t]
    \centering
    \includegraphics[width=\linewidth]{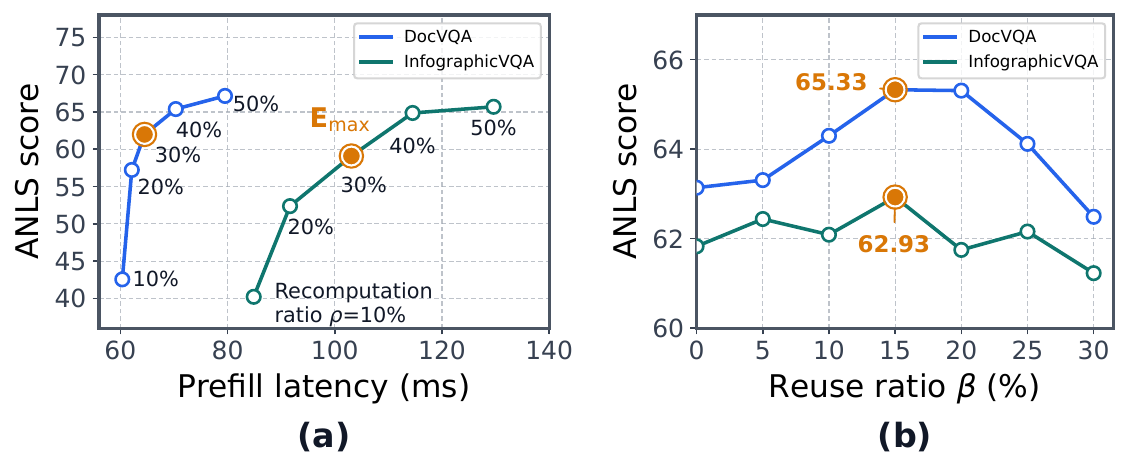}
    \caption{Sensitivity to hyperparameters.}
    \label{fig:recompute_tradeoff_pareto}
\end{figure}

\subsection{Sensitivity Analysis}\label{4.5}
In this section, we analyze the sensitivity of our system to two key hyperparameters, namely the \emph{approximate recomputation ratio $\rho$} and the \emph{reuse ratio $\beta$}, on the DocVQA and InfographicVQA datasets. The results are shown in~\autoref{fig:recompute_tradeoff_pareto}.

\noindent\textbf{Recomputation Ratio}. 
\autoref{fig:recompute_tradeoff_pareto}(a) shows the empirical Pareto trade-off among the approximate recomputation ratio, ANLS score, and prefill latency. Increasing the recomputation ratio consistently improves answer quality, but incurs progressively higher prefill overhead. Beyond 30\%, quality improvements become marginal and no longer offset the latency overhead. Accordingly, the normalized efficiency $E$ peaks at a 30\% recomputation ratio, indicating the best balance between output quality and system efficiency.

\noindent\textbf{Reuse Ratio}. 
\autoref{fig:recompute_tradeoff_pareto}(b) evaluates the impact of the reuse ratio on the ANLS score. The scores exhibit a \emph{non-monotonic trend}: increasing the reuse ratio initially improves performance, with the best results observed at round 15\%\textemdash 65.33 on DocVQA and 62.93 on InfographicVQA. This improvement indicates that reusing a moderate number of low-entropy tokens can effectively provide beneficial cached information. Beyond this point, the scores decline as increasing the reuse ratio introduces progressively more relatively high-entropy tokens, which add unexpected noise and degrade prediction quality. Overall, \system\ can still maintain high ANLS quality (above 61\%) at reuse ratios below 30\%.

\section{Related Work}
\noindent\textbf{Visual KV Cache Compression.}
Prior VLM inference optimizations primarily reduce visual prefill computation within each request. These works prune visual tokens according to redundancy or attention, while TokenMerging~\cite{bolya2023tome} and FrameFusion~\cite{fu2025framefusion} combine similar representations to shorten the visual sequence~\cite{yang2025visionzip,zhang2025sparsevlm,arif2025hired,chen2024fastv}. VL-Cache~\cite{tu2024vlcachecompression} and LOOK-M~\cite{wan2024lookm} instead reduce the memory cost of the visual KV cache. These methods reduce the prefill computation or KV cache memory, whereas \system\ extends the visual KV cache reuse across requests with approximate images.

\noindent\textbf{Visual KV Cache Reuse.}
Existing systems avoid repeated prefill via prefix caching~\cite{gim2024promptcache,ye2024chunkattention,zheng2024sglang}. Later works of position-independent caching relax prefix restriction for textual and multimodal inputs~\cite{hu2024epic,yao2025cacheblend,qin2025vlcache,ma2026kamera}, but visual reuse still requires identical visual matching. \system\ extends reusing to similar caches via patch reuse and recomputation.

\noindent\textbf{Hierarchical KV Cache Management.}
Existing memory management systems improve the memory utilization and access efficiency of LLM serving~\cite{kwon2023pagedattention,yang2025kvshare,liu2024cachegen,qin2025mooncake}, while approximate visual reuse introduces a different access pattern. Approximate hits require sparse KV cache reads and exhibit weaker temporal locality than exact hits, motivating \system's differentiated cache management.

\section{Conclusion}
This paper presented \system, an efficient serving system for approximate KV cache reuse for VLM inference. Motivated by the visual redundancy missed by exact matching, \system\ combined an entropy-aware KV cache reuse algorithm with differentiated hierarchical cache management. Our results showed that approximate caching improved cache hit rates and reduced prefill latency while preserving output quality. We view \system\ as a step toward broader KV cache reuse in VLM serving at scale. As workloads grow, new opportunities will emerge to exploit shared visual content across requests.

\newpage
\bibliographystyle{ACM-Reference-Format}
\bibliography{references}

\end{document}

%% file: tables/accuracy.tex
\begin{table*}[t]
\centering
\small
\begin{threeparttable}
\caption{Overall output-quality results using Qwen3-VL-8B~\cite{bai2025qwen3vl}.}
\label{tab:accuracy}

\setlength{\tabcolsep}{3.75pt}
\begin{tabular}{llccccccccc}
\toprule
& & \multicolumn{3}{c}{VLCache}
& \multicolumn{3}{c}{CacheBlend}
& \multicolumn{3}{c}{\system} \\
\cmidrule(lr){3-5}
\cmidrule(lr){6-8}
\cmidrule(lr){9-11}
Dataset
& Metric
& Overall
& Exact Hit
& Approx Hit
& Overall
& Exact Hit
& Approx Hit
& Overall
& Exact Hit
& Approx Hit \\
\midrule
VisionArena~\cite{chou2025visionarena}
& ANLS
& 73.10 & 77.39 & 40.17
& 84.83 & 75.66 & 41.29
& 85.02 & 76.79 & 57.43 \\
InfographicVQA~\cite{mathew2022infographicvqa}
& ANLS
& 60.04 & 41.42 & 31.35
& 60.95 & 40.14 & 39.37
& 62.09 & 40.74 & 59.09 \\
DocVQA~\cite{mathew2021docvqa}
& ANLS
& 52.66 & 56.62 & 44.94
& 44.28 & 55.23 & 32.94
& 64.30 & 56.11 & 62.21 \\
MMBench~\cite{liu2024mmbench}
& Accuracy
& 80.34 & 80.27 & 67.83
& 81.65 & 79.78 & 82.38
& 81.88 & 80.03 & 83.74 \\
SEED-Bench~\cite{li2023seedbench}
& Accuracy
& 59.59 & 54.40 & 53.24
& 58.48 & 54.25 & 49.24
& 60.74 & 54.35 & 58.57 \\
\midrule
\multicolumn{2}{l}{Average}
& 65.15 & 62.02 & 47.51
& 66.04 & 61.01 & 49.04
& 70.81 & 61.60 & 64.21 \\
\bottomrule
\end{tabular}
\end{threeparttable}
\end{table*}